\documentclass[notitlepage,10pt]{revtex4-1}
\usepackage{amsmath,amsfonts}
\usepackage{amssymb,enumerate}
\usepackage{graphicx}
\usepackage{amsthm}
\usepackage{latexsym}
\usepackage{graphicx}
\usepackage{hyperref}

\newtheorem{theorem}{Theorem}
\newtheorem{lemma}{Lemma}

\newtheorem{defn}{Definition}

\newcommand{\be}{\begin{equation}}

\newcommand{\ee}{\end{equation}}
\newcommand{\eenn}{\end{equation*}}
\newcommand{\benn}{\begin{equation*}}
\newcommand{\bea}{\begin{eqnarray}}
\newcommand{\eea}{\end{eqnarray}}
\newcommand{\bra}[1]{\left\langle #1\right|}
\newcommand{\ket}[1]{\left|#1\right\rangle}
\newcommand{\braket}[2]{\left\langle #1,#2\right\rangle}
\newcommand{\ketbra}[2]{|#1\rangle\langle #2|}
\newcommand{\pure}[1]{\ketbra{#1}{#1}}
\newcommand{\Tr}{\mathop{\mathrm{Tr}}}
\providecommand{\one}{\leavevmode\hbox{\small1\kern-3.8pt\normalsize1}}

\newcommand{\Z}{\mathbb{Z}}

\newcommand{\supp}{\mathrm{supp}}

\newcommand{\D}{\mathcal{D}}
\newcommand{\N}{\mathbb{N}}
\newcommand{\UB}{U_s(\partial A(2R))}
\newcommand{\WB}{W_s{(\partial A(2R))}}
\newcommand{\WBD}{W^\dagger_s{(\partial A(2R))}}
\newcommand{\UI}{U_s(A)}
\newcommand{\UE}{U_s(A^c)}
\newcommand{\UBD}{U^{\dagger}_s(\partial A(2R))}

\newcommand{\etal}{{\it et al. }}

\begin{document}

\title{Stability of the Area Law for the Entropy of Entanglement}
\author{S. Michalakis}
\affiliation{California Institute of Technology, Institute for Quantum Information and Matter}
\email{spiros@caltech.edu}
\begin{abstract}
Recent results \cite{tqo_stability,short_stability,michalakis:2011} on the stability of the spectral gap under general perturbations for frustration-free Hamiltonians, have motivated the following question: Does the entanglement entropy of quantum states that are connected to states satisfying an area law along gapped Hamiltonian paths, also satisfy an area law? We answer this question in the affirmative, combining recent advances in quasi-adiabatic evolution and Lieb-Robinson bounds with ideas from the proof of the $1$D area law \cite{1d_area_law}.
\end{abstract}

\maketitle
\section{Introduction}
Over the past decade, there has been a rapidly growing interest in the role of entanglement as a measure of complexity in simulating properties of quantum many-body systems~\cite{simulation_area_law, survey_area_law}. From a theoretical point of view, the question of how the entropy of entanglement scales with system size, has generated some spectacular results, such as Hasting's area law for one-dimensional gapped systems~\cite{1d_area_law} (see also \cite{michalakis:thesis, gs_approx} for partial generalizations), with a stronger bound for groundstates of frustration-free Hamiltonians given by Arad \etal in~\cite{new_area_law}, as well as the recent breakthrough by Brandao \etal in ~\cite{exp_decay_area_law}, where it is shown that exponential decay of correlations implies an area law for the entanglement entropy of one-dimensional states.

Here, we prove that the amount of entanglement contained in states connected along gapped, local Hamiltonian paths satisfies a certain stability property. Specifically, we show that groundstates of gapped Hamiltonians whose entanglement spectrum decays fast enough to imply an area law for the entanglement entropy, are connected via gapped, local Hamiltonian paths to states that satisfy a similar area-law bound. The techniques used to prove this result apply equally well to showing the stability of the area-law for any eigenstate of a Hamiltonian with fast enough decay in the entanglement spectrum, as long as the connection is along a gapped, quasi-local Hamiltonian path. As a consequence, groundstates of gapped, {\it frustration-free} Hamiltonians that satisfy the conditions for {\it stability} of the spectral gap under weak perturbations, found in \cite{tqo_stability,short_stability,michalakis:2011}, become central objects in the study of entanglement scaling in $2$D and in higher dimensions. This follows from the fact that the condition of {\it local indistinguishability} necessary for the stability of the spectral gap, also implies a rapid decay in the entanglement spectrum of the gapped groundstates \cite{michalakis:2011}. 

The main result appears in Theorem~\ref{thm:area-law}, which makes use of Lemma~\ref{lem:ent_bound} to convert a bound on the decay rate of the entanglement spectrum into a bound for the entanglement entropy of the states we are interested in. The main technical tool appears in Lemma~\ref{unitary_decomposition}, where we show that the unitary evolution of the initial groundstate along a gapped path can be approximated, up to rapidly-decaying error, with a product of three unitaries, two of which act on complementary regions and the third acts along the boundary separating the two regions. We expect that such a decomposition will find a variety of further uses, as it provides a more rigorous view of adiabatic evolution as a quantum circuit of local unitaries with depth dictated by the accuracy of the approximation we are willing to tolerate.

\section{Quasi-adiabatic evolution for gapped Hamiltonians}
We begin by introducing the central technical tool used in bounding the spread of the entanglement as  the initial state is adiabatically transformed. The main idea is to simulate the true adiabatic evolution with another evolution that satisfies two important properties:
\begin{enumerate}
\item The simulated evolution, introduced in~\cite{hast-quasi-intro} as {\it quasi-adiabatic continuation} and further developed in~\cite{tjo,quantum_hall,hast-quasi,BMNS:2011}, is indistinguishable from the true adiabatic evolution, if we restrict our attention to the evolution of uniformly gapped eigenspaces.
\item Unlike the true adiabatic evolution, the quasi-adiabatic evolution is generated by quasi-local interactions and, hence, transforms local operators into quasi-local operators.
\end{enumerate}
First, let us define precisely which families of Hamiltonians we are considering.
\begin{defn}\label{defn:path}
Let $H(s) = H_0 + \sum_{u\in \Lambda} V_u(s)$, where $\Lambda \subset \Z^d$ and set $b_u(r) = \{ v\in \Lambda: d(u,v) \le r\}$ to be the ball of radius $r$, centered on site $u\in \Lambda$. Then, the following assumptions hold:
\begin{enumerate}[i.]
\item $H_0 = \sum_{u \in \Lambda} Q_u$, with $Q_u$ supported on $b_u(r_0)$, for $r_0 \ge 0$ and $\|Q_u\|\le J_1$, for $u \in \Lambda$.
\item  $V_u(s)$ has support on $b_u(r_0)$, with $V_u(0) = 0$ and $\|V_u(s)\| \le J_1, \, \|\partial_s V_u(s)\| \le J_2$, for $s \in [0,1]$ and $u\in \Lambda$. 
\item $H(s)$ has spectral gap $\gamma(s) \ge \gamma > 0$, with differentiable groundstate $\ket{\psi_0(s)}$, for all $s \in [0,1]$. In particular, $H_0$ has unique groundstate $\ket{\psi_0(0)} := \ket{\psi_0}$, with spectral gap $\gamma(0) \ge \gamma$.
\end{enumerate}
\end{defn}
For the above one-parameter family of Hamiltonians $H(s)$, there exists a family of unitaries $U_s$ \cite{hast-quasi,BMNS:2011}, satisfying:
\be
\label{def:adiabatic_unitary} \ket{\psi_0(s)} = U_s \, \ket{\psi_0}, \quad
\partial_{s} U_s = i \, \D_s\, U_s, \quad U_0=\one, \quad \forall s \in [0,1],
\ee
where the generating dynamics, $\D_s$, has the following decay properties:
\be
\label{def:adiabatic_generator} \D_s = \sum_{u\in \Lambda} \sum_{r\ge r_0} \D_s(u;r),\quad \supp(\D_s(u;r)) = b_u(r), \quad \|D_s(u;r)\| \le 2\, J_2 \, f_{\gamma}(r-r_0),
\ee 
for a sub-exponentially decaying function $f_{\gamma}(r)$ (e.g. $\exp\{- c_0\, r/\ln^2r\}$, for $c_0 > 0$), with $f_{\gamma}(0)=1$ and decay rate proportional to $\gamma/v_0$, where $v_0 \sim J_1 \, r_0^{d-1}$ is the Lieb-Robinson velocity for the family of Hamiltonians $\{H(s)\}_{s\ge 0}$~\cite{hast-koma,lr3}.

\section{Decomposing the quasi-adiabatic evolution}
The above decay estimates on the generator of the unitary $U_s$, allow us to decompose the quasi-adiabatic evolution into a product of three unitaries, two of which act on disjoint subsets of the lattice and a third one coupling the disjoint evolutions at the boundary of the two sets. More importantly, the error in approximating the true adiabatic evolution decays sub-exponentially in the thickness of the boundary chosen for the coupling unitary.
\begin{lemma}\label{unitary_decomposition}
Let $U_s$ denote the unitary corresponding to the quasi-adiabatic evolution of a one-parameter, gapped family of Hamiltonians $H(s)$, as defined above. For $A \subset \Lambda$, set:
\begin{enumerate}
\item $I_{A}(R) = \{x \in A: d(x,\partial A) \le R\}$, 
\item $E_{A}(R) = \{x \in A^c: d(x,\partial A) \le R\}$, and 
\item $\partial A(R) = I_A(R) \cup E_A(R)$. 
\end{enumerate}
Then, there exist unitaries $\UI$,  $\UE$ and $\UB$ with non-trivial support on $A, A^c$ and $\partial A(2R)$, respectively, such that the following bound holds:
\be\label{bnd:decomp_error}
\|\UI \otimes \UE \, \UB - U_s\| \le \epsilon_s(R) := c_1 \Big(e^{c_2 (J_2/\gamma) |s|}-1\Big)\, |\partial A|\, f_{\gamma}(c_3 R),
\ee
for dimensional constants $c_1,c_2,c_3 >0$.
\begin{proof}
We sketch the proof, which uses the Lieb-Robinson bounds developed in \cite{hast-quasi,BMNS:2011}. First, define the following unitary operator:
\be\label{unitary:V}
V_{s}(A) = U^\dagger_s \, \UI \otimes \UE,
\ee
with $\partial_s U_s(A) = i \D_s(A) U_s(A), \, U_A(0) = \one$ and $\partial_s U_s(A^c) = i \D_s(A^c) U_s(A^c), \, U_{A^c}(0) = \one$, where we define, for $Z \subset \Lambda$: 
\be \D_s(Z) = \sum_{u\in \Lambda,\,r\ge 0: \,b_u(r)\subset Z} \D_s(u;r),
\ee
noting that $\supp(\D_s(Z)) \subset Z$ and, hence, $\supp(U_s(X)) \subset X$. Now, we set:
\be 
{\mathcal F}_s(\partial A) \equiv \D_s - (\D_s(A)+\D_s(A^c)) = \sum_{b_u(r) \cap \partial A \neq \emptyset} \D_s(u;r),
\ee 
where $X \cap \partial A \neq \emptyset \Longleftrightarrow (X \cap A \neq \emptyset) \wedge (X \cap A^c \neq \emptyset)$. The operator ${\mathcal F}_s(\partial A)$ can be well approximated by the locally supported $\D_s(\partial A(R))$. To see this, note that:
\be
{\mathcal F}_s(\partial A) - \D_s(\partial A(R)) = \sum_{u\in \Lambda} \sum_{r \ge d(u)} \D_s(u;r),
\ee
where $d(u) = \max\{R+1-d(u,\partial A), 1+d(u,\partial A)\}$. We may partition $\Lambda$ as follows:
\be
\Lambda = \cup_{k\ge 0} B_k, \quad B_k = \{u\in \Lambda: d(u,\partial A) = k\},
\ee
noting, further, that $|B_k| \le \sum_{u\in \partial A} (|b_u(k)|-|b_u(k-1)|) \le c_d k^{d-1} |\partial A|$, for $k\ge 1$ and $|B_0| = |\partial A|$, where $c_d = 2d$ can be thought of as the area of the unit ball in $\Z^d$.
This implies that:
\begin{align}
&\|{\mathcal F}_s(\partial A) - \D_s(\partial A(R))\| \le \sum_{k\ge 0} \sum_{u \in B_k} \sum_{r \ge d(u)} \|\D_s(u;r)\| \\
&\le 2\,J_2 \left(\sum_{k\ge 0}^{\lfloor (R+1)/2 \rfloor} |B_k| \sum_{r \ge \lceil (R+1)/2\rceil} f_{\gamma}(r-r_0) + \sum_{k\ge 1+\lfloor (R+1)/2 \rfloor} |B_k| \sum_{r \ge 1+k} f_{\gamma}(r-r_0)\right)\\
&\le 2\,J_2 |\partial A| \Big(\left(1+(R+1)^d\right) F_{\gamma}(\lceil(R+1)/2\rceil -r_0) + c_d \sum_{k\ge 1+\lfloor (R+1)/2 \rfloor} k^{d-1} F_{\gamma}(1+k-r_0)\Big),
\end{align}
where $F_{\gamma}(s) :=  \sum_{r \ge s} f_{\gamma}(r)$. 

At this point, given the decay rate of $f_{\gamma}$ \cite{hast-quasi,BMNS:2011}, it should be clear that for dimensional constants $c_0,d_0 > 0$:
\be
\label{bnd:gen}
\|{\mathcal F}_s(\partial A) - \D_s(\partial A(R))\| \le c_0\,(J_2/\gamma) |\partial A| (1+R)^{d+d_0} f_{\gamma}(\lceil (R+1)/2 \rceil - r_0).
\ee
By differentiating both sides of (\ref{unitary:V}), we get:
$
\partial_s V_{s}(A) = -\imath \left(U^\dagger_s \,{\mathcal F}_s(\partial A)\, U_s \right) V_{s}(A),\quad V_0(A)=\one.
$
We may approximate the unitary $V_s(A)$ with $W_s(A)$ generated by:
$
\partial_s W_{s}(A) = -\imath\, \left(U^\dagger_s \,\D_s(\partial A(R))\, U_s \right) W_{s}(A), \quad W_0(A) =\one,
$
such that:
\be\label{bnd:general}
\|V_s(A) - W_s(A)\| = \|W^\dagger_s(A)V_s(A) - \one\| \le \int_0^s \|\partial_t W^\dagger_t(A)V_t(A)\|\, dt \le |s| \, \sup_{t\in[0,s]} \|{\mathcal F}_t(\partial A)-\D_t(\partial A(R))\|.
\ee
Combined with (\ref{bnd:gen}), the above bound implies:
\be\label{bnd:unitary}
\|V_s(A) - W_s(A)\| \le c_0\,(J_2/\gamma)\, |s|\, |\partial A| (1+R)^{d+d_0} f_{\gamma}(\lceil (R+1)/2 \rceil - r_0), \quad c_0,d_0 > 0.
\ee
Finally, we approximate $W_s(A)$ with the unitary $\WB$ given by:
\be
\partial_s \WB = \imath \left(\UBD \,\D_s(\partial A(R))\, \UB \right) \WB, \quad W_0(\partial A(R)) = \one,
\ee
with the unitary $\UB$ generated by $\D_s(\partial A(2R))$.
Following (\ref{bnd:general}), we have:
\be\label{bnd:boundary}
\|W_s(A)-\WB\| \le |s| \sup_{t\in [0,s]} \| U^\dagger_t\,\D_t(\partial A(R))\, U_t - U^{\dagger}_t(\partial A(2R)) \,\D_t(\partial A(R))\, U_t(\partial A(2R))\|.
\ee
At this point, we cannot use the Lieb-Robinson bounds developed in \cite{hast-quasi},\cite[Thm. 4.6]{BMNS:2011} directly, since we are dealing with evolutions of the form $U^\dagger_s\, O_A \, U_s$ instead of $U_s\, O_A\, U^\dagger_s$. Nevertheless, by setting:
\be
F_t(O_X) = U^\dagger_t\,O_X\, U_t - U^{\dagger}_t(\partial A(2R)) \,O_X\, U_t(\partial A(2R)),
\ee
for an operator $O_X$ with support on $X \subset \Lambda$, we get:
\be
F_t(O_X) = \int_0^t \left(U^\dagger_s\, i[O_X,\D_s-\D_s(\partial A(2R))] \, U_s +F_s\left(i[O_X, \D_s(\partial A(2R))]\right) \right) ds,
\ee
which implies:
\be
\|F_t(O_X)\| \le |t| \sum_{b_u(r) \subsetneq \partial A(2R)} \sup_{s\in [0,t]} \|[O_X,\D_s(u;r)]\| + \sum_{b_u(r) \subset \partial A(2R)}  \sup_{s\in [0,t]} \|[O_X, \D_s(u;r)]\| \int_0^t \|F_s(O_X(u;r))\| \, ds,
\ee
where $O_X(u;r) = [O_X, \D_s(u;r)]/\|[O_X, \D_s(u;r)]\|$. At this point, we may use the recursive argument found in \cite{hast-koma,lr3}, setting $O_X = \D_t(\partial A(R))$ and recalling the sub-exponential decay of $\|\D_s(u;r)\|$. Setting: 
\be
\epsilon_s(R) := c_1 \left(e^{c_2 (J_2/\gamma) |s|}-1\right)\, |\partial A|\, f_{\gamma}(c_3 R),
\ee
for dimensional constants $c_1,c_2,c_3 >0$, we get from (\ref{bnd:unitary}) and (\ref{bnd:boundary}):
\be
\|U_s - \UI\UE\WBD\| \le \|V^\dagger_s(A)-W^\dagger_s(A) \| + \|W^\dagger_s(A) - \WBD \| \le \epsilon_s(R).
\ee
Renaming $\UB$ as $\WBD$ completes the proof.
\end{proof}
\end{lemma}

\section{Bounding the Entanglement Entropy:} At this point, we are ready to apply Lemma~\ref{unitary_decomposition} to study the entanglement entropy of states adiabatically connected to an initial state $\ket{\psi_0}$, whose Schmidt coefficients across a cut $A:A^c$ satisfy a certain rapid-decay condition. In particular, we assume that for a rapidly-decaying function $f_A(\cdot)$, with $f_A(0)=1$ and $A$ a convex subset of $\Lambda$:
\be\label{init_decay}
\sum_{\alpha \ge N^{R |\partial A|} + 1} \sigma_0(\alpha) \le f_A(R), \quad R \in \N,
\ee
where $N$ is the maximum single-site dimension and $\ket{\psi_0} = \sum_{\alpha \ge 1}
\sqrt{\sigma_0(\alpha)} \ket{\psi_{{A},0}(\alpha)}\otimes \ket{\psi_{{A^c},0}(\alpha)},$ the Schmidt decomposition of $\ket{\psi_0}$ across the cut $A:A^c$, with Schmidt coefficients in decreasing order.

The proof follows an argument similar to the one found in \cite{1d_area_law}.
First, we approximate the initial state $\ket{\psi_0}$ with the family of states $\{\ket{\psi_{0,R}}\}_{R\ge 0}$, where:
$$
\ket{\psi_{0,R}} = \frac{1}{\sqrt{c_R}} \sum_{\alpha = 1}^{N^{R\,|\partial A|}}
\sqrt{\sigma_0(\alpha)} \ket{\psi_{{A},0}(\alpha)}\otimes \ket{\psi_{{A^c},0}(\alpha)}
$$
and $c_R = \sum_{\alpha \ge 1}^{N^{R |\partial A|}} \sigma_0(\alpha) \ge 1-f_A(R)$, where we used (\ref{init_decay}).

The next step is to construct states with bounded Schmidt rank and increasing overlap to the adiabatically evolved state $\ket{\psi_0(s)} = U_s\ket{\psi_0}$. To accomplish this, we define the family of states: $$\ket{\psi_{0,R}(s)} \equiv \UI \otimes \UE \, \UB \ket{\psi_{0,R}}.$$
Setting the overlap $P(R) := |\braket{\psi_0(s)}{\psi_{0,R}(s)}|^2$ to be, say, at least $1/2$, fixes the minimum boundary thickness $R_0$ we consider in our approximation. Then, we use Lemma~\ref{lem:ent_bound} to show that the entanglement entropy of $\ket{\psi_0(s)}$ satisfies a non-trivial bound for all $s \in [0,1]$.
%THEOREM%
\begin{theorem}\label{thm:area-law}
For the gapped family of Hamiltonians $H(s)$ of Definition~\ref{defn:path}, let $\ket{\psi_0(s)}$ denote the groundstate of $H(s)$. Then, if the groundstate $\ket{\psi_0(0)}$ satisfies the decay condition (\ref{init_decay}), the entropy of $\rho_s(A) = \Tr_{A^c} \pure{\psi_0(s)}$ is bounded as follows:
\be\label{bnd:entanglement}
S(\rho_s(A)) \le 5\, (1+c_1)\, R_0\, |\partial A| + h_1, \quad R_0 = \min\{R\in \N: f_A(R_0) + 2 \epsilon_s(R_0) \le 1/2\},
\ee
where $c_1 = \sum_{n\ge 1} n \delta(n)$ and $h_1 = - \sum_{n\ge 0} \delta(n) \, \ln \delta(n)$, with $$\delta(n) = \big(f_A(n R_0)- f_A((n+1) R_0)\big) + 2 \big(\epsilon_s(n R_0)-\epsilon_s((n+1) R_0)\big)$$ and $\epsilon_s(\cdot)$ defined in \ref{bnd:decomp_error}.
\end{theorem}
Note that $c_1$ is a constant, as long as $\delta(n)$ decays faster than $n^{-(2+\epsilon)}$. Moreover, $h_1 = - \sum_{n\ge 1} \delta(n) \, \ln \delta(n) \le c_1 \ln t$, for $\delta(n) \ge t^{- n},\, n \ge 1,\, t > 1$ and $h_1 \le 1/\ln t$ if $\delta(n) \le t^{-n}$ for $n\ge 1$ and $t > 1$.

\begin{proof} Fix $s \in [0,1]$ and let $\rho_s(A) = \Tr_{A^c} \pure{\psi_0(s)}$.
Since we will be using the Schmidt decomposition of the state $\ket{\psi_0(s)}$, let us introduce it here:
$$\ket{\psi_0(s)} = \sum_{\alpha \ge 1}
\sqrt{\sigma_s(\alpha)} \ket{\psi_{{A},s}(\alpha)}\otimes \ket{\psi_{{A^c},s}(\alpha)},$$ where $\sum_{\alpha} \sigma_s(\alpha) = 1$ and $\{\ket{\psi_{A,s}(\alpha)}\}$, $\{\ket{\psi_{A^c,s}(\alpha)}\}$ form orthonormal sets supported on $A$ and $A^c$, respectively. We order the Schmidt coefficients of $\ket{\psi_0(s)}$ in decreasing order such that $\alpha<\beta \implies \sigma_s(\alpha) \geq \sigma_s(\beta)$. 
Moreover, we note that:
\be \rho_s(A) = \sum_{\alpha \ge 1} \sigma_s(\alpha) \pure{\psi_{A,s}(\alpha)} \qquad \mbox{ and } \qquad S(\rho_s(A)) = - \sum_{\alpha \ge 1} \sigma_s(\alpha) \ln \sigma_s(\alpha).
\ee
%$$\rho_{A^c} = \sum_\beta \sigma_0(\beta) \pure{\psi_{A^c,0}(\beta)}.$$
We define:
\be\label{overlap}
P(R) \equiv  |\braket{\psi_0(s)}{\psi_{0,R}(s)}|^2, \quad \Delta_s(R) = \UI \otimes \UE \, \UB - U_s
\ee
and use Lemma~\ref{unitary_decomposition} in order to relate $P(R)$ with the sub-exponentially decaying error $\epsilon_s(R)$:
\be\label{P:small}
P(R) = |\bra{\psi_0}\one + U^\dagger_s\Delta_s(R)\ket{\psi_{0,R}}|^2 \ge |\braket{\psi_0}{\psi_{0,R}}|^2 -2\|\Delta_s(R)\| \ge 1-(f_A(R)+2\epsilon_s(R)),
\ee
noting that $\braket{\psi_0}{\psi_{0,R}} = \sqrt{c_R}$.
Setting $R_0$ to be the smallest $R$ such that $f_A(R)+2\epsilon_s(R) \le 1/2$, the decay of $f_A(R)$ and $\epsilon_s(R)$ implies:
\be\label{P:large}
R \ge R_0 \implies P(R) \ge 1/2.
\ee
Now, the state $\ket{\psi_{0,R}(s)}$ has Schmidt rank bounded by $k_R \cdot N^{4\,R \, |\partial A|}$, where $k_R = N^{R\, |\partial A|}$ is the Schmidt rank of $\ket{\psi_{0,R}}$. To see this, first note that the Schmidt rank of $\UB \ket{\psi_{{A},0}(\alpha)}\otimes \ket{\psi_{{A^c},0}(\alpha)}$ is the same as that of $\UI\otimes\UE \,\UB \ket{\psi_{{A},0}(\alpha)}\otimes \ket{\psi_{{A^c},0}(\alpha)}$, along the boundary of $A$. It remains to see how the action of $\UB$ on $\ket{\psi_{{A},0}(\alpha)}\otimes \ket{\psi_{{A^c},0}(\alpha)}$ affects the Schmidt rank.

Since $\UB$ is an operator acting non-trivially only on sites in a subset of $\partial A(2R)$, we have the following general decomposition:
\begin{equation}\label{decomp}
\UB = \sum_{\alpha,\beta=1}^{D_I} \one_{A\setminus I_{A}(2R)} \otimes E(\alpha,\beta) \otimes G(\alpha,\beta) \otimes \one_{(A^c) \setminus E_{A}(2R)},
\end{equation}
where the matrices $G(\alpha,\beta)$ act on sites in $E_{A}(2R)$ and the matrix units $E(\alpha,\beta)$, which act non-trivially on $I_{A}(2R)$, form an orthonormal basis for $D_I \times D_I$ matrices. Moreover, $D_I \le N^{|I_{A}(2R)|} \le N^{2R |\partial A|}$, for convex $A$ and $N$ the maximum dimension among the single-site state spaces (e.g. $N=2$ for a system of qubits).

To bound the Schmidt rank of $$\ket{\psi_{B,0}(\gamma)} := \UB \ket{\psi_{{A},0}(\gamma)}\otimes \ket{\psi_{{A^c},0}(\gamma)},$$ we trace over sites in $A$ and study the rank of the operator:
\be\label{Schmidt_bound}
{\Tr}_{A} \left(\pure{\psi_{B,0}(\gamma)}\right) =
\sum_{\alpha=1}^{D_I}\left[ \sum_{\beta,\beta'=1}^{D_I}
c(\beta,\beta')\, \ketbra{F_{\alpha}(\beta,\gamma)}{F_{\alpha}(\beta',\gamma)}\right],
\ee
with 
\benn
\ket{F_{\alpha}(\beta,\gamma)}  = G(\alpha,\beta)\otimes \one_{A^c) \setminus E_{A}(2R)} \ket{\psi_{A^c,0}(\gamma)}
\eenn 
and 
\benn
c(\beta,\beta') = \bra{\psi_{A,0}(\gamma)} \one_{A\setminus I_{A}(2R)} \otimes E(\beta',\beta) \ket{\psi_{A,0}(\gamma)},
\eenn
coming from (\ref{decomp}).

Clearly, as a sum of $D_I$ matrices each with rank at most $D_I$, the above operator has rank bounded by $D_I^2$ and, hence, the Schmidt rank of $\UB\ket{\psi_A(\gamma,\delta)}$ is bounded above by $N^{4 R \,|\partial A|}$. Now, the final step is to note that if $\ket{\psi_{0,R}}$ has Schmidt rank $k_R$, then there are exactly $k_R$ states $\ket{\psi_{B,0}(\gamma)}$, each with Schmidt rank at most $N^{4 R \,|\partial A|}$, and the bound for the Schmidt rank of the approximation $\ket{\psi_{0,R}(s)} = \frac{1}{\sqrt{c_R}}\sum_{\gamma = 1}^{k_R} \sqrt{\sigma_0(\gamma)} \, \UI\otimes\UE\ket{\psi_{B,0}(\gamma)}$ follows.

The next step is to relate the overlap $P(R)$ to the Schmidt coefficients of $\ket{\psi_0(s)}$.
More specifically, we will now show that for $R \ge 0$:
\be
\label{constraint1}
\sum_{\alpha \le N^{5 R |\partial A|}} \sigma_s(\alpha) \geq P(R).
\ee
To prove this, first introduce the Schmidt decomposition of $\ket{\psi_{0,R}(s)}$: 
\be
\ket{\psi_{0,R}(s)} =\sum_{\beta = 1}^r \sqrt{\tau_{\beta}} \ket{\Phi_{A}(\beta)} \otimes \ket{\Phi_{A^c}(\beta)}, \mbox{ with } \sum_{\beta=1}^r \tau_{\beta} = 1 \mbox{ and } r \le N^{5R|\partial A|},
\ee 
as we have already demonstrated. For notational convenience, let 
$M_A(\alpha,\beta) = \left|\braket{\Phi_{A}(\beta)}{\psi_{{A},s}(\alpha)}\right|$ and $M_{A^c}(\alpha,\beta) = \left|\braket{\Phi_{A^c}(\beta)}{\psi_{{A^c},s}(\alpha)}\right|.$
Note that since $\{\ket{\Phi_{A}(\beta)}\}$, $\{\ket{\Phi_{A^c}(\beta)}\}$, 
$\{\ket{\psi_{A,s}(\alpha)}\}$ and $\{\ket{\psi_{A^c,s}(\alpha)}\}$ are orthonormal sets, Bessel's inequality implies that $$\sum_{\beta=1}^r M_A(\alpha,\beta)^2 \le 1 \mbox{ and } \, \sum_{\alpha} M_{A^c}(\alpha,\beta)^2 \le 1,$$ as well as 
$$\sum_{\alpha} M_A(\alpha,\beta)^2 \le 1 \implies \sum_{\alpha, \beta} M_A(\alpha,\beta)^2 \le r.$$

Then, an application of the triangle inequality followed by Cauchy-Schwarz gives the following upper bound for $P(R) = |\braket{\psi_{0,R}(s)}{\psi_{0}(s)}|^2:$
\begin{eqnarray*}
|\braket{\psi_{0,R}(s)}{\psi_{0}(s)}|^2 &\le&\left(\sum_{\alpha,\beta} \sqrt{\sigma_s(\alpha)}\sqrt{\tau(\beta)}\, M_A(\alpha,\beta) M_{A^c}(\alpha,\beta) \right)^2 \\
&\le&\left(\sum_{\alpha,\beta}\sigma_s(\alpha) M_A(\alpha,\beta)^2 \right) \, \left(\sum_{\alpha,\beta} \tau(\beta) M_{A^c}(\alpha,\beta)^2 \right) \\
&\le&\sum_{\alpha \le r} \sigma_s(\alpha).
\end{eqnarray*}
The last inequality follows from Schur convexity of $f([p(\alpha)]) = \sum_{\alpha} \sigma_s(\alpha) \, p(\alpha)$ and the observation that the vector $[1,1,\ldots,1,0,\ldots,0]$, with at most $r$ ones, majorizes $\left[\sum_{\beta} M_A(1,\beta)^2,\sum_{\beta} M_A(2,\beta)^2,\ldots\right]$. 

To see that $f([p(\alpha)])$ is Schur convex, note that if we set
$S_p(\alpha) = \sum_{k=1}^{\alpha} p(k)$ and $\Delta(\alpha,\beta) = \sigma_s(\alpha)-\sigma_s(\beta)$ then the condition that $[p(\alpha)]$ majorizes $[q(\alpha)]$ ($p \succeq q$) becomes $p \succeq q \Leftrightarrow S_p(\alpha) \ge S_q(\alpha),\, \forall \alpha$. Moreover, 
$$f(\{p(\alpha)\})-f(\{q(\alpha)\}) = \sum_{\alpha} \Delta(\alpha,\alpha+1) \left(S_p(\alpha)-S_q(\alpha)\right),$$
which is non-negative since we have arranged the $\sigma_s(\alpha)$ in decreasing order so that
$\Delta(\alpha,\alpha+1) \ge 0, \, \forall \alpha$.

Now that we have demonstrated (\ref{constraint1}), we may use it in combination with (\ref{P:large}) to show that $S(\rho_s(A))$ satisfies an entropy bound.
Using (\ref{P:small}-\ref{P:large}) and (\ref{constraint1}), we have for $n \ge 1$:
\be
\label{constraint2}
\sum_{\alpha\geq N^{5nR_0 \, |\partial A|}+1} \sigma_s(\alpha) \leq 1-P(nR_0) \le f_A(nR_0) + 2 \epsilon_s(nR_0).
\ee
We now maximize the entropy
$$S(\rho_s(A))=-\sum_{\alpha=1} \sigma_s(\alpha) \ln(\sigma_s(\alpha))$$
subject to the constraint (\ref{constraint2}).
Following the notation of Lemma~\ref{lem:ent_bound}, set $s_n = N^{5\,nR_0|\partial A|}, \, n \ge 1$ and $f(n) = f_A(nR_0)+2\epsilon_s(nR_0),\, n \ge 1$ and $f(0) = 1$, noting that $f(1) < f(0)$, by our choice of $R_0$. Then, for $r = N^{5R_0\,|\partial A|}$ and $s_1 = N^{5\,R_0|\partial A|}$, Lemma~\ref{lem:ent_bound} implies that:
\begin{equation}\label{CaseI:bound}
S(\rho_s(A)) \le 5 (1+c_1) R_0 \, |\partial A| + h_1,
\end{equation}
where $c_1 = \sum_{n\ge 1} n \delta(n)$ and $h_1 = - \sum_{n\ge 0} \delta(n) \, \ln \delta(n)$, with $$\delta(n) = (f_A(nR_0)-f_A((n+1)R_0)) + 2 (\epsilon_s(nR_0)-\epsilon_s((n+1)R_0)).$$
\end{proof}
We turn, now, to the following Lemma, which relates the decay properties of the entanglement spectrum to a bound on the entropy of entanglement:
\begin{lemma}[{\bf Entropy bound}]\label{lem:ent_bound} For $\rho$ a density matrix, let $\sum_{\alpha \ge 1} \sigma(\alpha) \pure{\psi(\alpha)}$ be its spectral decomposition, with $\sigma(\alpha)$ in decreasing order.  Assume that there is an increasing sequence of integers $\{s_n\}_{n\ge 0}$ with $s_n < s_{n+1} \le r\,s_n$, for $n\ge 1$ and $s_1 > s_0 = 0$, such that the following constraint holds for a strictly-decaying function $f(n)$, with $f(0)=1$:
\be
\sum_{\alpha \ge s_n + 1} \sigma(\alpha) \le f(n), \quad n \ge 0.\label{constraint}
\ee
Then, the entropy of $\rho$, given by $S(\rho) = -\sum_{\alpha \ge 1} \sigma(\alpha) \ln \sigma(\alpha)$, satisfies the following bound:
$$S(\rho) \le \ln s_1 + c_1 \ln r +h_1,$$
where $c_1 = \sum_{n\ge 1} n \delta(n)$ and $h_1 = - \sum_{n\ge 0} \delta(n) \, \ln \delta(n)$, with $\delta(n) = f(n)-f(n+1)$.
\begin{proof}
Since the Shannon entropy is Schur-concave, the von Neumann entropy $S(\rho)$ is bounded above by the Shannon entropy $H(\overrightarrow{\mu})$ of any probability distribution $\{\mu(\alpha)\}_{\alpha\ge 1}$ consistent with the constraint (\ref{constraint}), that is majorized by $\{\sigma(\alpha)\}_{\alpha\ge 1}$ (i.e. $\overrightarrow{\mu} \preceq \overrightarrow{\sigma}$.)

Define $\delta(n) = f(n)-f(n+1), \, n \ge 0$. For $1\le \alpha \le s_1$, the constraint $\sum_{\alpha \ge 1} \sigma(\alpha) = 1$ implies:
$$\sum_{\alpha=1}^{s_1} \sigma(\alpha) \ge f(0)-f(1) \implies \sum_{\alpha=1}^{s_1} \mu(\alpha) = \delta(0) \implies \mu(\alpha) = \frac{\delta(0)}{s_1}, \quad 1\le \alpha \le s_1.$$
Similarly, for $\alpha \in [s_n+1, s_{n+1}], \, n \ge 1,$ we see that we should choose $\overrightarrow{\mu}$ to satisfy:
\benn
\sum_{\alpha=s_n+1}^{s_{n+1}} \mu(\alpha)= \, f(n)-f(n+1) \implies \mu(\alpha) = \frac{\delta(n)}{s_{n+1}-s_n}.
\eenn
Gathering terms, we decompose the Shannon entropy $H(\overrightarrow{\mu})$ as $\sum_{n = 0}^{\infty} H_n(\overrightarrow{\mu})$, where we have defined $H_n(\overrightarrow{\mu}) = - \sum_{\alpha = s_n+1}^ {s_{n+1}} \mu(\alpha) \ln \mu(\alpha),\, n\ge 0$. We can bound each $H_n(\overrightarrow{\mu})$ as follows, recalling that $s_{n+1} \le s_1 r^n$ and, hence, $s_{n+1} - s_n \le s_1 r^n$:
\begin{equation}
H_n(\overrightarrow{\mu}) \le - \delta(n) \, \ln \delta(n) + n \delta(n) \ln r + \delta(n) \ln s_1, \quad n\ge 0.
\end{equation}
Using $\sum_{n\ge 0} \delta(n) =1$, we get the bound:
\begin{eqnarray*}
H(\overrightarrow{\mu}) &\le& \ln s_1 + c_1 \ln r + h_1,
\end{eqnarray*}
where we defined $c_1 = \sum_{n\ge 1} n \, \delta(n)$ and $h_1 = - \sum_{n\ge 0} \delta(n) \, \ln \delta(n)$. This completes the proof of the Lemma.
\end{proof}
\end{lemma}
\section{Conclusion}
We have shown that states satisfying an area law are connected via gapped, local Hamiltonian paths to states that satisfy a similar bound on their entropy of entanglement. This result holds for lattice systems in any dimension. Since the above result relies only on Lieb-Robinson bounds and locality properties of the {\it quasi-adiabatic evolution}, it can be shown that even gapped paths with long-range interactions satisfy the bound~(\ref{bnd:entanglement}), as long as the decay is fast enough (i.e. power-law decay with a dimension-dependent exponent.) We hope this work will motivate interested readers to work out the details of this argument, which would have important implications for the classification of phases, given the connection of states satisfying an area-law to Matrix Product States (MPS) in $1$D and Projected Entangled Pair States (PEPS) in $2$D~\cite{class_phases_1d, schuch_class_phases}. 

\noindent {\bf Acknowledgements:}
The author would like to thank M. B. Hastings for helpful remarks on the use of quasi-adiabatic evolution in the context of entanglement entropy, as well as fruitful discussions with B. Nachtergaele, N. Schuch and A. Gorshkov.

\end{document}